\documentclass[a4paper,12pt,oneside,article]{article}

\usepackage{hyperref}
\usepackage[utf8]{inputenc} 
\usepackage[T1]{fontenc} 
\usepackage{lmodern}
\usepackage{amsmath, amssymb, bm, mathtools, mathdots}
\usepackage{lineno}

\usepackage[normalem]{ulem}
\usepackage{array, booktabs, tabularx}
\usepackage{graphicx, caption, subfig, xcolor}
\usepackage[retainorgcmds]{IEEEtrantools}
\usepackage{cancel}
\usepackage{cite}
\usepackage{authblk}

\graphicspath{ {./images/} }
\usepackage{enumerate}
\def\rm{\mathrm}

\def\cm{\;\rm{cm}}

\def\m{\; \rm{m}}
\def\kW{\;\rm{kW}}
\def\ms{\; \rm{m/s}}
\def\bn{\begin{enumerate}}
\def\en{\end{enumerate}}
\def\be{\begin{equation}}
\def\ee{\end{equation}}
\def\bea{\begin{eqnarray}}
\def\eea{\end{eqnarray}}
\def\ba{\begin{array}}
\def\ea{\end{array}}

\def\bi{\begin{itemize}}
\def\ei{\end{itemize}}

\author[1,2]{Maria Alice Gasparini}
\affil[1]{\small Institute of Teacher Education, University of Geneva, Switzerland}
\affil[2]{\small Departement of Physics, University of Geneva, Switzerland}

\begin{document}

\title{What Happens if You Put Your Head in the Geneva Water Jet? An Inquiry-Based Physics Activity Exploring Fluid Dynamics}

\maketitle

\begin{abstract}
We describe a physics education activity for third-year Bachelor students, inspired by a humorous question about the Geneva water jet. The exercise engages students in key scientific practices: reformulating everyday questions in scientific terms, constructing simplified models, performing semi-quantitative estimations, and comparing alternative solution methods. Students explore approaches based on Bernoulli’s principle and a power analysis, revealing consistent results when assumptions are carefully considered. The activity emphasizes critical reasoning, including identifying relevant data, making approximations, and applying energy and mass conservation to incompressible fluids. It also fosters metacognitive skills and higher-order thinking (HOT), illustrating the universality of fundamental physical principles across diverse phenomena. By situating the task in a relatable, real-world context, the activity motivates students while exposing them to problem-solving challenges rarely encountered in traditional instruction, such as Fermi-type estimation and cross-context knowledge transfer.

\end{abstract}

\section{Introduction}\label{intro}

The Geneva water jet provides an ideal context for exercises or activities to be proposed to secondary school students, particularly in the teaching of mechanics.
Teachers commonly use it to address uniformly accelerated rectilinear motion (UARM) by studying the free fall of a small water droplet within the jet. In this situation, the droplet can be considered to be ``shielded'' from air resistance, meaning that the latter does not significantly affect its motion.\\

Using data provided by the {\it Services Industriels de Genève} (SIG) and presented in the Tab. \ref{Tab:tab1} students can, for example, be asked to estimate the theoretical height of the jet. By simply using the exit velocity of the water, one obtains a value of approximately 
$h=160 \;\rm{m}$. This result can then serve as a basis for discussing the validity of the assumption that air resistance can be neglected \cite{cours2DF}. 
\begin{small}
\begin{center}
\captionof{table}{Technical specifications of the Geneva water jet from \cite{SIG, lombard}.}
{\renewcommand{\arraystretch}{1.5}
\renewcommand{\tabcolsep}{0.2cm}
\begin{tabular}{|p{4cm}|p{4cm}|p{4cm}|} 
		\hline 
		{{\bf Quantity}} & {\bf Formula} & {\bf Value (SIG)}    \\ 
		\hline 		\hline 
		{Maximum jet height} & { $h$} & { 140 m}  \\ 
		\hline 
		{Water exit speed} & {$v$} & {200 km/h}  \\ 
		\hline 		
		{Water flow rate} & {$dV/dt$} & {500 L/s}  \\ 
		\hline 		
		{Mechanical power} & {$P$} & {1 MW}  \\ 
		\hline 		 
		{Lighting power} & {$P_L$} & {9 kW} \\ 
		\hline 
	\end{tabular} } \label{Tab:tab1}
\end{center}
\end{small}

The same result -- still neglecting air resistance -- can subsequently be obtained by applying the principle of {\it conservation of mechanical energy} to the motion of a droplet (see, for example, \cite{lombard}, or \cite{cours3DF}, or \cite{web1}). This second approach demonstrates that different methods can lead to the same result.\\

Presenting these two approaches offers two important pedagogical advantages. First, although it is conceptually simpler for students to use the kinematic equations of uniformly accelerated motion (two equations involving time), this method remains technically more demanding for them. The conservation-of-energy approach -- although more abstract -- illustrates, in this context, its simplicity and effectiveness. Second, students can observe how a plurality of problem-solving approaches helps to deepen their understanding of a physical phenomenon and to appreciate the generalising strength of physical laws and principles.\\

In the remainder of this article, we present an extension of this exercise by broadening it to include aspects of fluid dynamics, as implemented in the elective course {\it Physique du Quotidien} (``Physics of Everyday Life''), intended for third-year Bachelor’s students in physics at the University of Geneva.\\

Although the physical and mathematical concepts addressed in this extension do not exceed the level of the end of secondary education or the first years of a Bachelor’s program, the activity presents pedagogical interest in several respects. In particular, it illustrates how, starting from simple observations drawn from everyday life, it is possible to construct key steps of the scientific process and to raise students’ awareness of the nature of science (NOS) \cite{NOS}.
This begins with the formulation of scientifically relevant qualitative questions. These questions can then be addressed through modelling and semi-quantitative reasoning, allowing a range of skills to be developed: the ability to make approximations and assess their impact, the comparison of alternative problem-solving approaches, the evaluation of uncertainties, and the study of limiting cases as a method for checking the relevance of results, thereby contributing, among other things, to the development of metacognitive skills.\\

This approach also illustrates how new relevant questions often emerge during the course of the analysis, highlighting the open-ended and potentially unbounded nature of scientific inquiry. The educational value of this activity thus lies in the fact that it brings together a wide range of competencies within an authentic and motivating application context.

\section{Questioning and modelling the situation}\label{modelling}

\begin{figure}[!h]
	\centering
        	\includegraphics[width=0.6\textwidth]{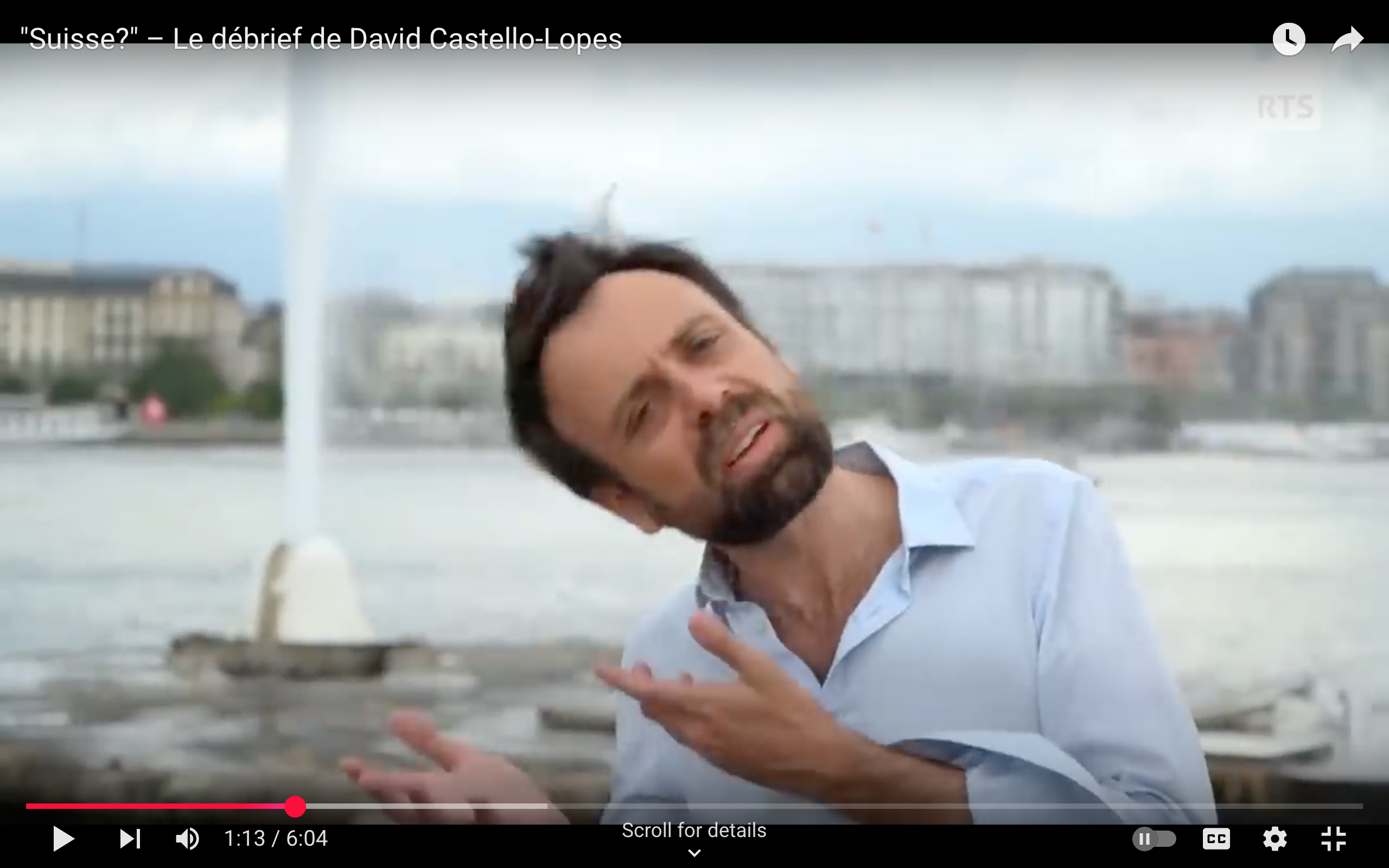}\\
	\caption{Journalist and comedian Castello-Lopes wonders about the effect of a water jet on a person's head \cite{lopes2}. 
    }
    \label{fig:DCL}
 \end{figure}

The journalist and humorist David Castello-Lopes became well known to the Swiss general public from 2020 onwards through his appearances on the {\it Radio Télévision Suisse} (RTS) programme {\it 52 minutes}, in which he presented a series of video segments addressing Swiss stereotypes \cite{lopes1}. In the episode broadcast on 22 May 2021, he reflects on the production of these segments and mentions the one devoted to the Geneva city, in which he refers to the water jet \cite{lopes2}. He explains that he then asked himself the following question (see Fig. \ref{fig:DCL}):

\begin{center}
``If one places one’s head at the base of the water jet, does one die?''
\end{center}

This question may appear amusing or naïve, but it constitutes an excellent entry point for a physicist, insofar as it can be reformulated as a scientific question and used as the starting point for a genuine inquiry-based process. For this reason, the first task proposed to the students consisted in reformulating this question in a scientifically meaningful way, using the data provided in Tab. \ref{Tab:tab1}.\\

Students generally reformulated the question in terms of water pressure or of the force exerted on a surface placed at the base of the jet (in this case, a person’s head). For the majority of them, force appeared to be the physical quantity most intuitively connected to their everyday experience. The question ultimately retained was the following:

\begin{center}
{\bf Q1}: ``Can one estimate the force exerted by the water jet on a head placed at its base?''
\end{center}

Developing such a question cannot proceed without recognising the need to simplify the modelling relative to the real situation: it is assumed that the person’s head can be modelled as a rigid, flat surface, located at the outlet of the jet, which brings the outgoing water to rest upon contact.

Such a simplification may appear crude, but within the scientific approach it is commonly the ``price to pay'' in order to pursue the investigation and to introduce physical quantities, laws, and quantitative estimates.

\section{(Semi-)quantitative estimation and mathematization}\label{maths}

While the scientific reformulation of the question generally does not pose major difficulties, the development of a {\it quantitative} model to answer it using only the information provided in the table of Tab. \ref{Tab:tab1} proved much more challenging for the majority of students. This difficulty was particularly evident in the absence of complete attempts -- even incorrect ones --despite around ten minutes of work and the opportunity to discuss with a peer.\\

This situation does not result from a lack of prior knowledge, as the necessary notions had already been covered in other courses. Numerous studies show that transfer failure is largely due to a combination of cognitive overload \cite{CLT1, CLT2}, deficits in metacognitive control  \cite{meta1, meta2} and insufficient explicit instruction of transfer as a specific skill \cite{trans1, trans2}. More specifically, students show limited familiarity with problems that are:

\begin{enumerate}

\item Located outside the scope of a given chapter or a specific teaching context -- in this case, mechanics or fluid dynamics.

\item Of the ``Fermi problem'' type \cite{fermi}, which notably involve:

\begin{itemize}

\item Selecting relevant information, which requires the ability to anticipate a modelling scheme (theory, laws, principles) from one’s entire body of knowledge. Such anticipation requires an organised mental representation that novices have not yet systematically developed \cite{CLT3, novices}.

\item Identifying redundant data among the information provided (for example, in the table of Tab. \ref{Tab:tab1}).

\item Estimating the order of magnitude of missing data, such as the contact area between the person’s head and the water.
\end{itemize}

\end{enumerate}

Combined with the approximations necessary for modelling, as discussed in the previous paragraph, these skills are often absent from traditional pedagogical approaches. This is especially true when they appear simultaneously and within an authentic context. Yet, these skills are intrinsically part of the nature and specificity of scientific research \cite{NOS}.

\section{Analysis of an approach: Bernoulli’s principle}\label{bernoulli}

\begin{figure}[!h]
	\centering
        	\includegraphics[width=\textwidth]{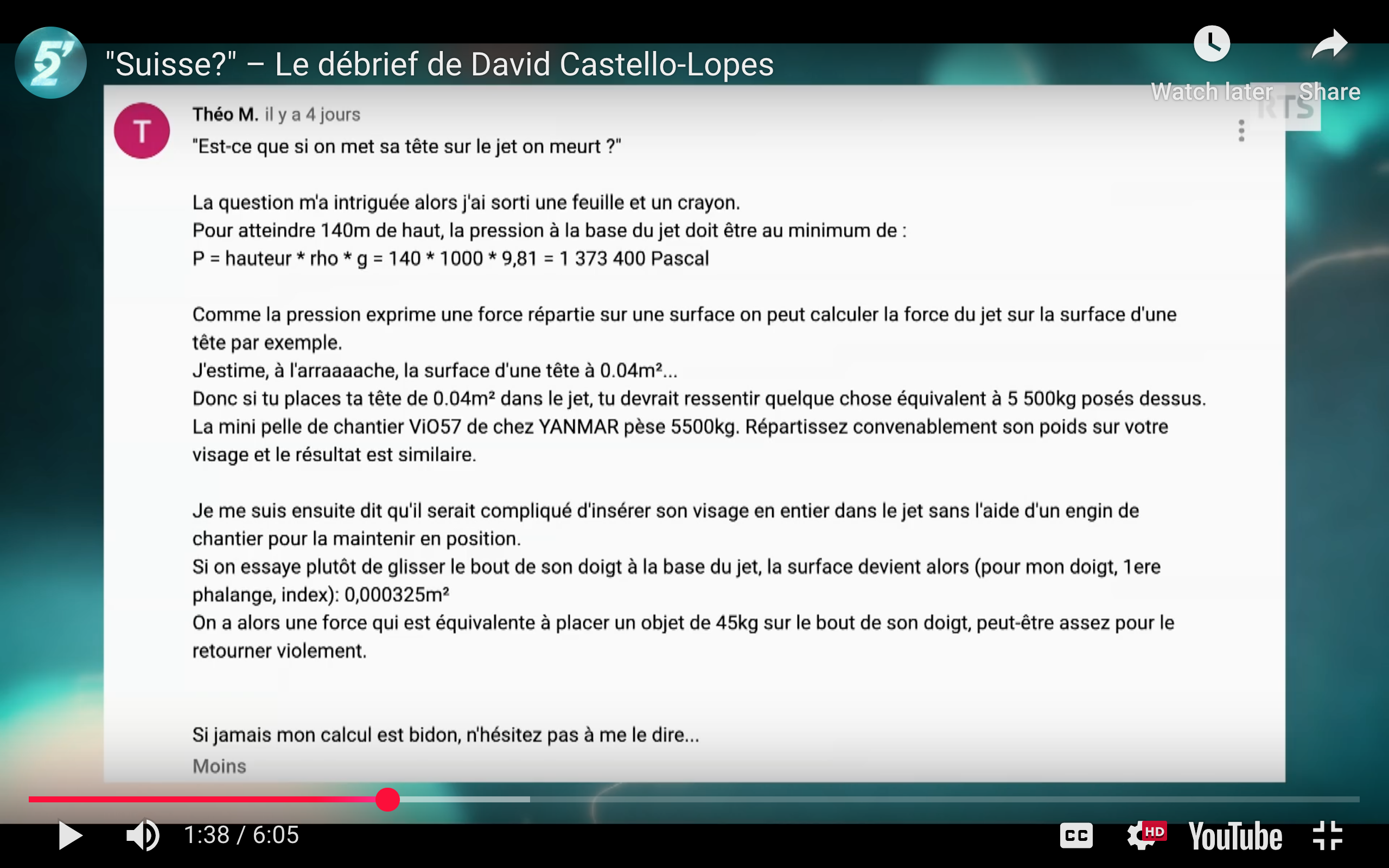}\\
	\caption{A solution proposed by an internet user \cite{lopes2}. 
    }
    \label{fig:internaute}
 \end{figure}

Prior to the broadcast, internet users were able to interact with the journalist David Castello-Lopes on social media, and one of them proposed a solution to the question posed, as shown in Fig. \ref{fig:internaute}. Once the students examined this solution, they found that it was well within their reach.

Although the Bernoulli principle employed by the internet user is a relevant and effective tool for addressing the question, the students immediately identified an inconsistency in the reasoning. The idea used actually relies on a generalisation of the conservation of energy principle applied to a {\it single body} (e.g., a droplet) to the conservation of mechanical energy {\it density} of a continuous {\it substance} (in this case, water). In this framework, pressure can be interpreted as energy per unit volume, and the pressure on the surface stopping the water jet at its base would correspond to the mechanical energy density of the water in the jet. If mechanical energy were conserved between the water’s exit and its maximum height, one could write:
\be
p=\frac{E_{\rm{mec}}}{V} = \frac{E_{\rm{kin}}(0)}{V} =\frac{E_{\rm{pot}}(h)}{V} = \rho g h
\ee

where $p$ is the pressure at the level of the head, $E_\rm{mec}$ and $V$ are the mechanical energy and the volume of a droplet, $\rho$ is the water density and $h$ is the height measured from the base of the jet.\\

However, calculating the height from the exit velocity (which gives $h=160$ m rather than 140 m) shows that a fraction of the mechanical energy is dissipated during the ascent of the jet. Consequently, if we wish to obtain a more accurate estimate of the pressure at head level, the correct solution is:
\be\label{pressure}
p= \frac{E_{\rm{kin}}(0)}{V} =\frac{1}{2}\,\rho\,v^2 = 0.5\cdot 1'000 \;\rm{kg/m^3}\cdot (55.6 \;\rm{m/s} )^2 =1.54\cdot 10^6 \;\rm{Pa}
\ee
\be\label{bern1}
\Rightarrow \quad F_{\rm{head}}=p\cdot S_{\rm{head}} = 1.54\cdot 10^6 \;\rm{Pa}\cdot 0.04\, \rm{m^2} \approx 62\, \rm{kN}\,.
\ee

This correction modifies the final value by approximately 10 to 20\%, without, however, changing the order of magnitude of the force obtained by the internet user.\\

Some students noted that the estimate of $0.04 \m^2$ for the contact area between the head and the jet seems excessive: a more realistic value would be on the order of $S_{\rm{head}}=0.02 \;\rm{m^2}$, giving:
\be\label{bern2}
F_{\rm{head}}=p\cdot S_{\rm{head}} = 1.54\cdot 10^6 \;\rm{Pa}\cdot 0.02\, \rm{m^2} \approx 31\, \rm{kN}\,.
\ee

Finally, students also suggested the possibility of alternative solution approaches based on the data provided by the SIG, shown in the table of Tab. \ref{Tab:tab1}. These approaches were assigned as homework, with the objective of obtaining the same order of magnitude for the force exerted.

\section{Power-based approach}\label{power}

Some students noted that it is also possible to determine the force exerted on the head using the power of the water jet provided in the table of Tab. \ref{Tab:tab1}: $P=1'000$ kW. When a stationary surface stops the water flow, the power absorbed by the surface can be related to the force $F$ exerted on it by the jet (assuming that other forces acting on the surface are negligible compared to that of the jet). Since power represents the rate of change of energy, we can write:
$$
P=\frac{dE_{\rm{kin}}}{dt}=\frac{1}{2}\, m\cdot 2v\cdot\frac{dv}{dt} = v\cdot F
$$
\be\label{pow}
\Rightarrow\quad F= \frac{P}{v} =\frac{1'000\kW}{55,6\ms} \approx 18\, \rm{kN}\,.
\ee

We note that:
\bn
\item Although of the same order of magnitude, the result obtained from Eq. \refeq{pow} is significantly smaller than that obtained using Bernoulli’s principle, regardless of the value chosen for 
$S_{head}$. This difference is too large to be explained solely by the imprecision of the data.

\item The value obtained does not necessarily correspond to the force exerted on the head, if the latter occupies only a fraction of the total cross-section of the water jet. To obtain the actual value of $F_{\rm{head}}$, the result of Eq. \refeq{pow} would then need to be corrected by multiplying it by the ratio $S_{\rm{head}}/S_{\rm{jet}}$, if this ratio is smaller than unity.
\en

These observations led to the following questions, assigned as homework:
\begin{center}
{\bf Q2}: ``How can this discrepancy be explained? Which approach is correct, and why?''
\end{center}

\section{Cross-section of the jet flow}\label{section}

The students predominantly adopted two different approaches:
\bi
\item A first group searched the web for the diameter of the jet, finding a value of approximately 16 cm, which gives $S_\rm{jet} \approx S_\rm{head} = 0.020\m^2$, comparable to the area of the section of a human head. This estimate therefore leads to the same result as Eq. \refeq{pow}, namely a force of about 18 kN on the head, and thus does not explain the difference observed between the two previous methods.

\item A second group, following the instruction to use only the data provided by the SIG in the table of Tab. \ref{Tab:tab1}, determined the cross-sectional area of the jet from the flow rate $dV/dt$:
\be
\frac{dV}{dt} = S_\rm{jet}\cdot v \quad \iff\quad S_\rm{jet} = \frac{dV/dt}{v}=\frac{0.5\,\rm{m^3s^{-1}}}{55.6\;\rm{m\,s^{-1}}} = 0.0090\m^2
\ee
This result shows that the surface area of the head is in fact smaller than that of the jet and does not cover the entire cross-section of the flow. Consequently, the relevant value to be used in Eqs. \eqref{bern1} and \eqref{bern2} is $S_\rm{jet} = 0.0090\m^2$, rather than $0.02\m^2$:
\be\label{bern3}
F_{\rm{head}}=p\cdot S_{\rm{jet}} = 1.54\cdot 10^6 \;\rm{Pa}\cdot 0.0090\, \rm{m^2} \approx 14\, \rm{kN}\,.
\ee

\ei

\subsection{Data consistency}

The approach based on the jet’s power thus highlights a second inaccuracy in the internet user’s solution, related to the estimation of the jet’s cross-sectional area. However, even after correction, the result of Eq. \refeq{pow}, obtained from the power, remains different from that of Eq. \refeq{bern3}, based on Bernoulli’s principle. 
This naturally leads to a new question:

\begin{center}
{\bf Q3}: ``How can one explain, despite this correction, a difference exceeding 20\% in the estimated force?''
\end{center}

Students' first try was to question the consistency of the available data. Since these are publicly accessible data intended for a general audience rather than scientific purposes, their precision can at best be assumed to be to one significant digit, or at the order-of-magnitude level. To obtain the result of Eq. \refeq{pow}, we used the jet power provided in the table of Tab. \ref{Tab:tab1}, $P=1'000\kW$. However, it is also possible to estimate this power from the other data in the same table. Considering a continuous water flow with a volumetric rate $dV/dt =0.5\,\rm{m^3/s}$ and a constant velocity $v=55.6\ms$ (the velocity variation over a short distance at the base being negligible), the kinetic energy transferred per unit time is:

\be\label{pownew}
P=\frac{dE_\rm{kin}}{dt}=\frac{1}{2}\, \rho \,\frac{dV}{dt}\cdot v^2 = 500 \cdot 0.5 \cdot 55.6^2 =773\kW\approx 800\kW
\ee

This estimate is of the same order of magnitude as the value reported by the {SIG}, but it still differs by more than 20\%. Moreover, if this new value (Eq. \eqref{pownew}) is used for the power in Eq. \refeq{pow}, the result obtained coincides exactly with that obtained using Bernoulli’s principle:

\be\label{pow1}
\Rightarrow\quad F_\rm{head}= \frac{P}{v} =\frac{773\kW}{55,6\ms} \approx 14\, \rm{kN}\,.
\ee

\subsection{Comparison of methods}

At this stage, it is natural to examine the connection between the two approaches. In the estimation of power from the other data in the table of Tab. \ref{Tab:tab1}, several intermediate substitutions were made without appearing explicitly in the final numerical result. To clarify this relationship, one simply needs to replace the symbolic expression for power obtained in Eq. \eqref{pownew} into \eqref{pow}:

\be\label{pow2}
F_\rm{head}= \frac{P}{v} =\frac{\frac{1}{2}\, \rho \,\frac{dV}{dt}\cdot v^2}{v} =\frac{1}{2}\, \rho\, \frac{dV}{dt}\cdot v\,.
\ee

The volumetric flow rate can be written as:

\be\label{flow}
\frac{dV}{dt} = S_\rm{jet}\cdot v \quad \Rightarrow\quad F_\rm{head}=\frac{1}{2}\, \rho \cdot S_\rm{jet}\cdot v^2 = p\cdot S_\rm{jet}\,.
\ee

By performing these substitutions, it becomes clear that the final expression for the force coincides exactly with that obtained from the application of Bernoulli’s principle. This equality demonstrates that both approaches rely on the same underlying physical principles, but that the Bernoulli-based approach provides the most direct and transparent way to establish the relationship between velocity, pressure, and force via the conservation of (the density of) mechanical energy.

\section{Development of student questions}\label{questions}

Two additional questions emerged during the students’ discussions, without the students being able to resolve them immediately.

\subsection{Jet diameter}

The group of students who looked up the jet diameter on the web ($d=16\cm$) quickly raised a first question:

\begin{center}
{\bf Q4}: ``How can we explain that the cross-sectional area of the jet derived from its diameter does not correspond to the actual cross-section of the water flow?''
\end{center}

\begin{figure}[!h]
	\centering
        	\includegraphics[width=0.6\textwidth]{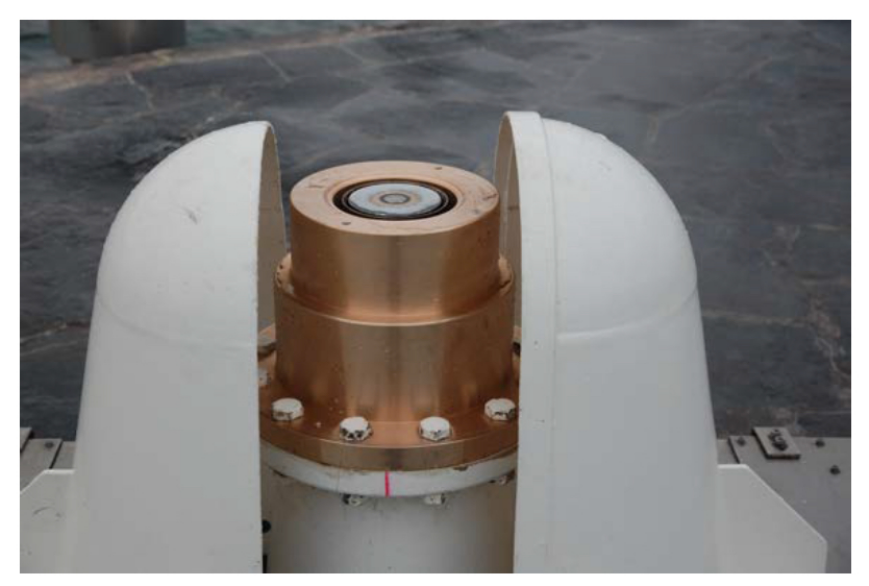}\\
	\caption{Geneva water jet base \cite{hees19}. 
    }
    \label{fig:buse}
 \end{figure}

The answer lies in the actual shape of the flow: its cross-section is not a solid disk but rather a ring (i.e., a hollow shape), as illustrated in Fig. \ref{fig:buse} \cite{hees19}. 

This observation could be used to directly determine the thickness of the ring, which is about 1 cm -- an exercise that we leave to the reader. We chose istead to guide the students toward a different question:

\begin{center}
{\bf Q5}: ``What is the relevance of calculating the force applied to the entire cross-section of the flow, knowing that the head’s surface, $S_\rm{head}$, which is circular and solid, cannot completely cover the annular section of the jet?''
\end{center}

It then appears more appropriate to express the result in terms of pressure, that is, force per unit area perpendicular to the surface, expressed in $\rm{N/cm^2}$. From Eq. \eqref{pressure}, we obtain:

\be
p=1.54\cdot 10^6 \,\rm{\frac{N}{m^2}} \approx 160  \,\rm{\frac{N}{cm^2}}\,,
\ee

corresponding to the pressure exerted by a weight of approximately 16 kg on a $1\cm^2$ surface.

\subsection{Relationship between water exit velocity and flow cross-section}

The second question arose when a student connected the problem to an everyday observation: when using a hose, if one partially blocks the opening with a finger, the water speed increases. The student then imagined that, similarly, a human head reducing the cross-section of the jet could increase the exit velocity.\\

In our simplified model, this reasoning does not hold: the head is located after the jet’s exit and does not alter the outgoing flow, but only slows the water by absorbing its kinetic energy.

However, this raises the following question:

\begin{center}
{\bf Q6}: ``Does the observation that the exit velocity depends on the cross-section of the flow contradict Bernoulli’s principle, which states that, without a change in potential energy, the exit velocity depends only on the pressure at the base (or in a garden hose)?''
\end{center}

A first answer can be obtained through a simple experiment: fill a plastic container (for example, a 5 L family-size detergent bottle) and drill several holes with different cross-sections at the same height. The range of the jets -- hence the exit velocity -- will be the same for all holes, and may even be shorter for the smaller ones due to friction. Moreover, the range decreases proportionally with the water height in the container, i.e., with the pressure.\\

In a system with a constant flow rate (such as the base of the water jet or a garden hose), in addition to energy conservation, we must consider mass conservation: because the fluid is incompressible, the inflow and outflow rates must be equal. If the exit cross-section were reduced, the pressure at the base of the jet would increase, which would cause the exit velocity to increase in order to maintain a constant flow rate. Thus, the velocity would increase, but the pressure would also increase, which does not contradict Bernoulli’s principle. Energy conservation therefore remains valid, but it must be considered together with mass conservation in this context.

\section{Conclusion}\label{conclusion}

Starting from a humorous question about the Geneva water jet, we developed an activity encompassing several essential steps of the scientific process: reformulating the question in scientific terms, modelling the situation, semi-quantitative mathematization, and comparing different solution methods.\\

Through successive questioning, students were able to observe that the different approaches -- whether applying Bernoulli’s principle or the jet-power method -- yield consistent results when all assumptions and corrections are properly considered. The discrepancies highlighted the limited precision of publicly available data and the importance of evaluating approximations. Discussions of the jet’s actual cross-section and the effect of the head on the flow illustrated the complementarity between energy and mass conservation in incompressible fluids.\\

Students were thus challenged to make approximations, identify relevant information from imperfect data, and engage their critical and metacognitive skills to assess the relevance of their results -- competencies that are rarely mobilized simultaneously in traditional teaching settings.\\

This activity aligns closely with the transversal objectives of the third-year Bachelor’s optional course, including:

\bi
\item Fostering curiosity and careful observation skills, characteristic of the scientific mindset. 
Even everyday situations that appear trivial -- or even whimsical, as in this case -- can raise significant scientific questions. Combined with serendipity, this ability to ``see'' the questions behind seemingly ordinary phenomena has historically led to many scientific advances. As Louis Pasteur observed, ``{\it In the field of observation, chance favours only the prepared mind}'' \cite{pasteur}.

\item Mobilising multiple elements of prior knowledge outside specific chapters to formulate and solve Fermi-type problems, emphasising coherence of reasoning over numerical precision, supporting knowledge organisation \cite{CLT1, CLT2, CLT3, novices}, and explicitly strengthening metacognitive capacities and higher-order thinking (HOT) skills in science \cite{Zoh15}.

\item Identifying fundamental concepts and principles underlying seemingly diverse phenomena, illustrating the universality of natural laws. 
Abstract conservation principles, in particular, demonstrate their effectiveness and simplicity: the generalisation of energy conservation to fluid motion -- the Bernoulli principle -- allows not only the analysis of the water jet’s dynamics but also applies to a wide range of domains, such as ventilation systems (natural or artificial), aviation, and hydraulic devices. As Feynman noted, ``{\it 
What we are looking for is how nature behaves. If you look at it carefully, you see that it is always the same -- it's always the same, only the numbers change.}'' \cite{feyn}.
\ei

Finally, although this analysis is not easily verified experimentally, more than two years after the humorous broadcast inspiring this activity, a tourist reportedly attempted to place his head in the jet -- an incident covered by local media. The attempt was not fatal, but upon leaving the hospital, he received a fine for crossing safety barriers \cite{20min}.

\bigskip
\listoffigures

\end{document}